\documentclass[preprint,amssymb,showpacs,superscriptaddress,floatfix]{revtex4-1}
\usepackage{amssymb}
\usepackage[percent]{overpic}
\usepackage{graphicx}
\usepackage{mwe}
\usepackage{braket}
\usepackage{mathtools, amssymb, amsmath}
\usepackage{cancel}
\usepackage{float}
\usepackage{mathrsfs}
\usepackage{gensymb}
\usepackage[utf8]{inputenc}
\usepackage{amsmath}
\usepackage{graphicx}
\usepackage{dcolumn}
\usepackage{bm}
\usepackage[title]{appendix}
\usepackage{subcaption}
\usepackage{bm}
\usepackage{natbib,hyperref}
\usepackage{hyperref}
\usepackage{color}
\usepackage[dvipsnames]{xcolor}
\usepackage[normalem]{ulem} 

\hypersetup{colorlinks=true, 
    linkcolor=red,          
    citecolor=magenta,        
    filecolor=gree,      
    urlcolor=blue           
}

\usepackage{float}

\usepackage[utf8]{inputenc}
\usepackage[T1]{fontenc}
\usepackage{mathptmx}
\usepackage{etoolbox}

\begin{document}
\title{Modeling Memristor-Based Neural Networks with Manhattan Update: Trade-offs in Learning Performance and Energy Consumption}

\author{Walter Quiñonez}
\affiliation{Laboratorio de Ablación Láser (INN-CONICET-CNEA), Gral. Paz 1499 (1650), Buenos Aires, Argentina.}
\affiliation{Universidad de Buenos Aires. Facultad de Ciencias Exactas y Naturales.  Departamento de F\'isica. C1428 Ciudad Autónoma de Buenos Aires, Buenos Aires, Argentina.}

\author {María José Sánchez} 
\affiliation{Centro At\'omico Bariloche and Instituto Balseiro (UNCuyo), 8400 San Carlos de Bariloche, R\'io Negro, Argentina.}
\affiliation{Instituto de Nanociencia y Nanotecnolog\'{\i} (INN-CONICET-CNEA), nodo Bariloche, Av. Bustillo s/n (8400), San Carlos de Bariloche, Argentina.}

\author{Diego Rubi*} 
\affiliation{Laboratorio de Ablación Láser (INN-CONICET-CNEA), Gral. Paz 1499 (1650), Buenos Aires, Argentina}

\email[Corresponding author:]{ diego.rubi@gmail.com}

\date{\today}

\begin{abstract}

We present a systematic study of memristor-based neural networks trained with the hardware-friendly Manhattan update rule, focusing on the trade-offs between learning performance and energy consumption. Using realistic models of potentiation/depression (P/D) curves, we evaluate the impact of nonlinearity (NLI), conductance range, and number of accessible levels on both a single perceptron (SP) and a deep neural network (DNN) trained on the MNIST dataset. Our results show that SPs tolerate P/D nonlinearity up to $\mathrm{NLI} \leq 10^{-2}$, while DNNs require stricter conditions of $\mathrm{NLI} \leq 10^{-3}$ to preserve accuracy. Increasing the number of discrete conductance states improves convergence, effectively acting as a finer learning rate. We further propose a strategy where one memristor of each differential pair is fixed, reducing redundant memristor conductance updates. This approach lowers training energy by nearly 50\% in DNN with little to no loss in accuracy. Our findings highlight the importance of device–algorithm co-design in enabling scalable, low-power neuromorphic hardware for edge AI applications.

\end{abstract}
\maketitle

\section{Introduction}


Artificial neural networks (ANNs) have become indispensable tools in modern computing \cite{Geron}, enabling breakthroughs in fields such as image recognition, autonomous systems, speech processing, and personalized recommendations \cite{LeCun2015, Bojarski2016, Hinton2012, Zhang2019}. These networks rely on a large number of interconnected computational nodes and adjustable weights to learn complex patterns from data. However, the hardware implementation of such networks remains constrained by the limitations of traditional von Neumann architectures, where the separation between memory and computation leads to significant inefficiencies due to frequent data transfers \cite{Mehonic2022, Zou2021}.

To address this "memory wall," alternative computing paradigms are being actively explored. Among them, memristor-based hardware offers a promising route by combining computation and memory storage within a single nanoscale device \cite{yu_2017,Mehonic2022, Zolfagharinejad2024, Mehonic2024}. Memristors are resistive switching devices whose conductance can be modulated by electrical pulses and retained without power \cite{saw_2008,iel_2016, Acevedo_2017,acevedo_2018,Acevedo_2020,Ferreyra2020}, allowing them to act as synaptic weights in hardware neural networks \cite{Alibart_2013,Prezioso_2014, Chen_2015, Li_2018, Joshi_2020, Schmidgall2024, Aguirre2024}. When arranged in crossbar arrays, memristors naturally enable in-memory vector-matrix multiplications (VMM) \cite{Mannocci2023, Yang2013, Ielmini2021} —the computational bottleneck in most ANN operations—while drastically reducing energy consumption and latency. However, the performance of ANNs is usually affected by the so-called \emph{non-idealities} of the memristive devices that constitute their building blocks. These non-idealities include non-linear potentiation–depression curves, a discrete number of accessible conductance levels within a limited  range defined by the $G_{\text{MAX}}$ and $G_{\text{MIN}}$ values, in addition to device-to-device and cycle-to-cycle variations \cite{Chen_2015, Burr_2015, Shibata_2020, Apsangi_2022, Joksas2022, Qui_onez2023, Glint2024, Yon2022, Gutsche2021, Bengel2021, Kim2022}.
While memristive systems have shown considerable promise in accelerating inference tasks \cite{Alibart_2013, Kim2021}, enabling online training remains a more demanding challenge \cite{Alibart_2013, Prezioso_2014, Li_2018}. Training directly on hardware requires robust and reliable weight update mechanisms that cope with the analog nature and physical non-idealities of memristor devices. In this work, we simulate the behavior of memristor-based perceptron and deep neural networks with in situ training, focusing specifically on how device-level characteristics affect network performance in terms of convergence behavior, classification accuracy, and energy efficiency.

The Manhattan update rule \cite{Zamanidoost_2015, Prezioso_2014}, appears as a simplified gradient-based approach particularly suited for hardware applications. This rule updates weights incrementally in fixed steps based on the sign of the gradient, making it well-suited for memristive devices that support discrete conductance levels and can be programmed with single pulses. Compared to more complex update schemes, the Manhattan rule reduces circuit overhead and increases robustness against device variability, while still maintaining learning performance in proof-of-concept simple networks such as perceptrons \cite{Alibart_2013,Prezioso_2014}

Despite its simplicity of implementation and inherent advantages, no systematic studies on the suitability of more complex ANNs using the Manhattan rule under realistic constraints for memristors have been reported in the literature. This paper aims to fill this gap by numerically investigating the influence of three key physical parameters on ANN performance under the Manhattan rule: 

i) \textbf{Linearity of potentiation and depression curves.} Ideal weight updates assume that memristor conductance changes are proportional to the number of programming pulses. In real memristors, however, the update curves are often highly nonlinear, which can distort learning dynamics and impede convergence; 

ii) \textbf{Limited conductivity windows (from $G_{\text{MIN}}$ to $G_{\text{MAX}}$}) The ratio between the maximum and minimum conductance values defines the dynamic range available for weight encoding. A narrow window can limit learning capacity or lead to premature saturation, while a wider window allows finer control but may increase sensitivity to noise; 

iii) \textbf{Multilevel conductance resolution.} The number of stable intermediate states in the memristor’s conductance directly affects the granularity -defined as the minimum resolvable change in conductance that the device can reliably achieve- of weight updates \cite{li2018}. Higher resolution can improve accuracy but also demands tighter control over device programming and variability.

By incorporating realistic device models into our simulations, we capture the impact of these non-idealities on ANN performance. We benchmark both single perceptrons and deep neural networks using the MNIST handwritten digits database under various parameter regimes to evaluate the trade-offs involved in \textit{in situ} learning. Additionally, we analyze the energy consumption associated with training, accounting for both read and write operations, to assess the practical viability of these systems for low-power edge computing applications.

This work provides a comprehensive view of the challenges and opportunities in implementing memristive neural networks with online training. Our study contributes to the co-design of efficient, adaptive neuromorphic hardware. The use of the Manhattan update rule offers a realistic and hardware-friendly pathway toward scalable and trainable memristive AI systems.

\section{Methods}

Artificial neural networks (ANNs) consist of interconnected layers of neurons, where the strength or synaptic weight ($w_{ij}$) of the connection between neurons is iteratively trained for a specific task. A standard architecture for ANNs in machine learning is the feedforward architecture, in which neurons in one layer connect only to neurons in the subsequent layer. This means there are no connections between neurons within the same layer or with previous layers, and no recurrent loops \cite{Haykin2009}. When an ANN consists of only one layer beyond the input layer, it is called a single-layer perceptron (SP). If it has one or more hidden layers, it is called a deep neural network (DNN). Following a basic training scheme for a DNN, a training dataset is fed into the network, and its performance is evaluated using an activation function. The synaptic weights are then adjusted by minimizing a loss function.

\begin{figure}[H]
\centering
\includegraphics[width=1 \textwidth]{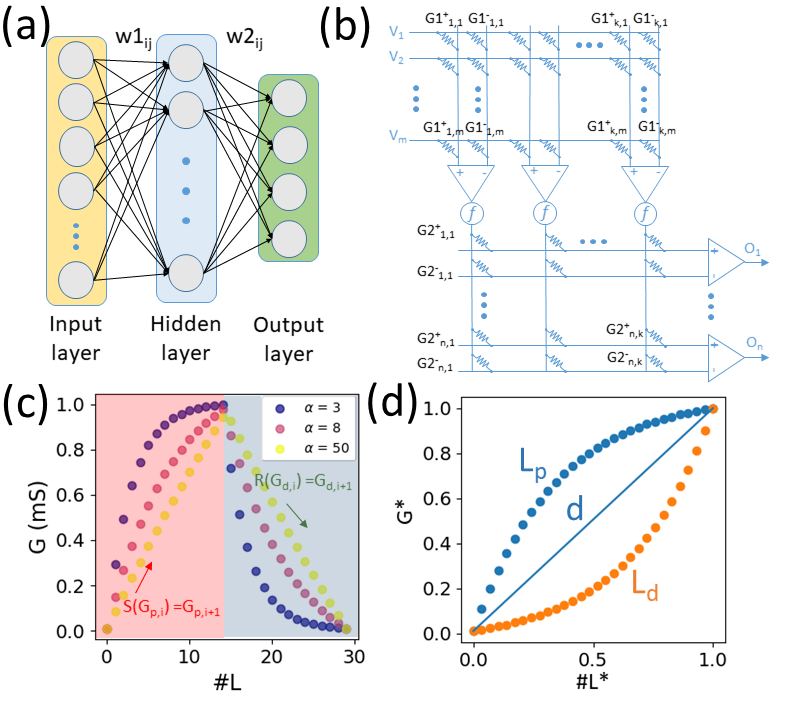}
\caption{(a) Schematic representation of an ANN with a single hidden layer. (b) Electrical scheme of a crossbar array architecture for the hardware implementation of the ANN shown in (a). (c) Synthetic potentiation (red region) and depression (blue region) curves obtained for different parameters $\alpha$ and fixed $\#$L. (d) Normalized P/D curves as a function of normalized L. To compute NLI, we consider d to be the length of the ideal (linear) segment connecting the first and last points of the P/D curves, while L$_p$ and L$_d$ represent the lengths of the potentiation and depression curves, respectively.} 
\label{Fig.1}
\end{figure}

We aim to simulate the on-line training of a DNN physically implemented on memristor crossbar arrays, where the conductances of memristors are linked  to the DNN synaptic weights. Fig. \ref{Fig.1}(a) displays a sketch of an ANN including a single hidden layer. The weights between the input and hidden layers are represented by the matrix elements $w1_{ij}$, while the weights between the hidden and output layers are represented by $w2_{ij}$. 


In Fig. \ref{Fig.1}(b), we present the electrical scheme of a crossbar array, which serves as the hardware implementation of the DNN shown in Fig. \ref{Fig.1}(a). Following the approach proposed by Prezioso et al. \cite{Prezioso_2014}, the synaptic weights of the ANN are encoded as the conductances of differential memristor pairs \cite{Alibart_2013}, such that $wl_{ij} = Gl^+_{ij} - Gl^-_{ij}$, with $l=1,2$. This scheme enables both positive and negative weights, thereby overcoming the intrinsic limitation that individual device conductances are strictly positive. The trade-off is a doubling of the required memristors, with the subtraction of currents physically realized by a differential amplifier. Following this scheme, the hidden layer of the DNN is represented by the conductance matrices $G1^+_{ij}$ and $G1^-_{ij}$, while the output layer is represented by $G2^+_{ij}$ and $G2^-_{ij}$. The output current $I^1_i$ of the hidden layer can be calculated using Ohm’s and Kirchhoff’s laws as 

\begin{equation}
I^1_i = \sum_{j=1}^{m} \left( G1^+_{ij} - G1^-_{ij} \right)V_j,      
\label{Eq1}
\end{equation}

where $V_j$ is the input data vector encoded as voltage values and we consider that the  layer  contains  m x 2 k devices, with j and i being the number of inputs and outputs, respectively. To compute the output i, $O_i^1$, an activation function $f$ is applied as $O_i^1 = f(\beta I^1_i)$, where $\beta$ is a scaling factor that depends on the conductance range of the devices and the voltage range used to encode the input data. The outputs of the hidden later feed the next layer of the crossbar array and a similar procedure produces the final outputs (currents) of the ANN. After the training of the network, the optimal values of conductance for each memristor in the array are known and it is possible to perform inference tasks. 

During the training procedure, the standard gradient descent (GD) method can be followed in order to compute the update of the synaptic weights $\Delta w_{ij}$. However, mapping these weight updates onto the conductance values of each differential pair is electrically challenging. A simpler alternative is the Manhattan rule \cite{Zamanidoost_2015, Prezioso_2014}, which is based on GD but enables conductance updates by applying fixed SET (increasing conductance) or RESET (decreasing conductance) voltage pulses at each training step. With this rule, we have a trade-off between learning precision and ease of implementation.
Specifically, the Manhattan rule  dictates: i) if $\Delta w_{ij} > 0$ a SET (RESET) voltage pulse is applied to modify $G^+_{ij} (G^-_{ij})$; ii) if $\Delta w_{ij} \leq 0$, a RESET (SET) voltage pulse is applied to modify $G^+_{ij} (G^-_{ij})$.

To update the conductance according to the Manhattan rule, we follow granular P/D curves, usually observed in memristors, synthetically generated according to $G_p = B\left(1 - \exp\{-n_i / \alpha\} \right) + G_{\min}$ for potentiation and $G_d = -B\left(1 - \exp\{(n_i - \#L) / \alpha\} \right) + G_{\max}$ for depression. In both cases, $n_i$ is the discrete variable that represents the different conductance levels in the P/D curves, $\alpha$ is a parameter that controls its linearity, $G_{\min}$ is the lowest value of the conductance range of the device and $\#L$ is the overall number of available levels in the conductance window. The parameter $B$ is defined as $B = (G_{\max} -G_{\min} ) /\left(1 - \exp\{-\#L / \alpha\} \right)$ , where $G_{\max}$ is the maximum conductance value. We will only consider symmetric synthetic curves, in the sense that both the potentiation and depression curves share the same number of levels $\#L$, conductance range, and parameter $\alpha$.  

In Fig. \ref{Fig.1}(c), we show synthetic P/D curves for $\#L = 15$ and $\alpha = 3, 8, 50$, where it can be seen that, for a fixed value of $\#L$, the parameter $\alpha$ controls the linearity of the curve, as larger values of $\alpha$ lead to  more linear curves. However, if both parameters are varied simultaneously the linearity results from their combined effect -as expected from the exponential dependence on  $\#L/\alpha$-.
To update the conductance values $G^+_{ij}$ and $G^-_{ij}$ according to the Manhattan rule, we define two discrete functions, S and R, which reproduce the evolution of the P/D curves as SET or RESET voltage pulses are applied, respectively.
To define these functions, we consider that the discrete values of the potentiation curve satisfy $G_{p,i}$ $\leq$ $G_{p,i+1}$, while those of the depression curve satisfy $G_{d,i}$ $\geq$ $G_{d,i+1}$. This is condensed in the ordered sequences $\{G_{p,i}\}_{i=1}^{\#L}$ and $\{G_{d,i}\}_{i=1}^{\#L}$. 
To capture this, we define $S(G_{p,i}) = G_{p,i+1}$ and $R(G_{d,i}) = G_{d,i+1}$. We also define $S(G_{p,\#L}) = G_{p,\#L}$ and $R(G_{d,\#L}) = G_{d,\#L}$ as a boundary condition, since in both cases $G_{p,\#L}$ and $G_{d,\#L}$ are the last values of each sequence. Lastly, since the learning rule might dictate a decrease(increase) in conductance even when the current value belongs to the potentiation(depression) curve, special care must be taken when applying updates to ensure consistency with the physical behavior of the device. For these cases we define $S(G_{d,i}) = \min_{j} \{\|G_{d,i} - G_{p,j}\| \  | \ G_{d,i} < G_{p,j} < G_{p,\#L} \}$ 
and $R(G_{p,i}) = \min_{j} \{\|G_{p,i} - G_{d,j} \| \ | \  G_{d,1} < G_{d,j} < G_{p,i}   \}$.   
We consider these functions to provide a good approximation of the general response of memristive devices to the accumulation of voltage pulses, while preserving both the granularity and the distribution of the points that define the potentiation/depression (P/D) curves.

Different metrics can be proposed to quantify the degree of non-linearity of P-D curves, such as mean absolute value (MAE), root-mean square error (RMSE) or maximum deviation \cite{Iida_2024}; however, each one has its own limitations: MAE may underestimate the impact of localized nonlinearities by averaging errors over the entire curve; RMSE is strongly influenced by outliers or noise, which can overshadow the general shape of the trajectory; and maximum deviation captures only the largest pointwise discrepancy, making it highly sensitive to measurement artifacts and unrepresentative of the overall curve behavior.


We propose, instead, a non-linearity index (NLI) computed as follows: 
(i) we normalize $\#L$ as $\#L^*$ and $G$ ($G_d^*$ and $G_p^*$) so that they range from zero to one within each curve (this normalization is performed only for the NLI calculation) ; 
(ii) we compute the length $d$ of the straight segment connecting the first and last points of the curve as $d=\sqrt{2}$; 
(iii) we calculate the lengths $L_p$ and $L_d$ along the potentiation and depression curves, respectively, as the sum of the distances between successive points: $L_{p/d}=\sum_{i=1}^{\#L^*-1}\sqrt{(n^*_{i+1} - n^*_i)² +\left(G^*_{({p/d})_{i+1}}-G^*_{({p/d})_{i}}\right)^2}$, where $n_i^*$ is the normalized discrete variable that represents the different conductance levels in the P/D curves; 
and (iv) we compute $\text{NLI}=(L_{p/d}-d)/d$. 
This procedure is schematically illustrated in Fig.~\ref{Fig.1}(d). 
Therefore if  $L_{p/d}\approx d$, the NLI approaches  zero. Otherwise, the NLI increases, as a signature of the greater non-linearity in the curve shape. 

The main advantage of the NLI is that it offers a simple, model-independent, and unit-free metric to quantify how much the potentiation or depression trajectories deviate from an ideal linear trend, compactly capturing the overall curvature of the conductance evolution in a single index.

\begin{figure}[H]
\centering
\includegraphics[width=1 \textwidth]{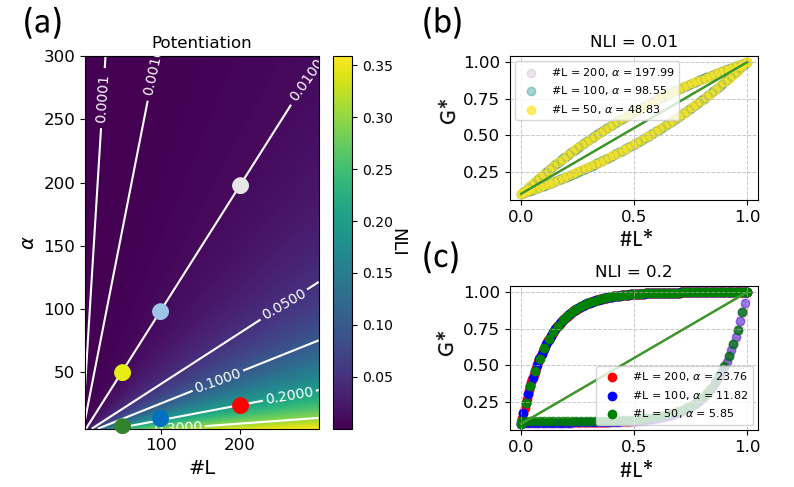}
\caption{(a) Heatmap corresponding to NLI as a function of parameters $\alpha$ and $\#L$. White lines represents contour lines of the function. Only the heatmap for the potentiation is presented (depression presents the same heatmap). The extracted P/D curves for each combination of parameters (colored points) on the contour line for NLI = 0.01 and NLI = 0.2 are shown in pannels (b) and (c) respectively, for normalized   $G^{*}$ and  $\#L^{*}$. For points on the same contour line, the shape of the normalized P/D curve are similar to each other. Heatmaps are similar across different conductance ranges.} 
\label{Fig.2}
\end{figure}

In Fig.~\ref{Fig.2}(a), we show a heatmap of the NLI as a function of the parameters $\alpha$ and $\#L$ for potentiation curves, with both parameters varying in the range 0--300. We note that the number of levels in typical P/D curves is usually on the order of tens, although some reports have demonstrated up to several hundred levels \cite{Xiao2025}, consistent with the range chosen for our simulations. Since we are considering symmetric P/D curves, the same heatmap applies to depression, and similar results are obtained for different conductance ranges (See
Figs. SM-1). White lines represent contour lines corresponding to constant NLI values, reflecting a linear relationship between $\#L$ and $\alpha$.
In Fig.~\ref{Fig.2}(b), we show extracted P/D curves for points along the contour line corresponding to $\text{NLI} = 0.01$: $\#L = 50$, $\alpha = 48.83$ (yellow); $\#L = 100$, $\alpha = 98.55$ (light blue); and $\#L = 200$, $\alpha = 197.99$ (gray). The normalized curves retain the same shape along the contour, indicating that $\#L$ can be increased without altering the curve shape, provided the parameters follow the linear relation defined by the contour. The same trend is observed in Fig.~\ref{Fig.2}(c) for $\text{NLI} = 0.2$: $\#L = 50$, $\alpha = 5.85$ (green); $\#L = 100$, $\alpha = 11.82$ (blue); and $\#L = 200$, $\alpha = 23.76$ (red).  
Because the NLI is computed on normalized P/D curves, the heatmaps are independent of the absolute conductance range, reflecting linearity properties that depend only on the curve shape.

To estimate the energy dissipated by the crossbar during one epoch of training, we consider two contributions to the total energy $E_t$: one from the forward pass of the input images through the crossbar, $E_{\text{f}}$, and the other from the energy dissipated by each device during the conductance update, $E_{\text{u}}$, such that $E_t = E_{\text{f}} + E_{\text{u}}$.
Assuming Joule heating as the primary dissipation mechanism, $E_{\text{f}}$ can be computed as
$E_{\text{f}} = \Delta t_{\text{f}} P_{\text{f}}$, where $\Delta t_{\text{f}} = 10$ ns, which is a typical target timescale for memristor writing and reading \cite{vonWitzleben2021, Jiang2016}. The forward dissipated power is computed as $P_{\text{f}} = \sum_{\mu} \sum_{i} \sum_{j} \left( V{^\mu_{j}} \right)^2 G_{ij}$.
Here, $\left( V^\mu_j \right)^2$ is the square of the $j$-th component of the $\mu$-th input vector, and $G_{ij}$ is the conductance matrix of the input layer. For a single perceptron (no hidden layers), the calculation is straightforward. However, for a DNN, only the energy dissipation in the input layer can be directly computed, since the output voltages of this layer are required to evaluate the next one.

Because we compute only the output currents of each layer, obtaining the corresponding output voltages would require solving a 2D resistor network for each input image at every epoch, which would make the simulations highly computationally expensive.
To circumvent this problem, we estimate the energy dissipation in the output layer of the DNN by computing the average energy dissipated per memristor in the input layer and multiplying it by the number of memristors in the output layer.

On the other hand, the energy dissipated during the conductance update of a specific memristor is given by the integral $\int_{t_0}^{t_f} V_{\text{u}}^2(t)G_{ij}(t)dt$, where $V_{\text{u}}(t)$ is the applied voltage pulse (SET or RESET), $G_{ij}(t)$ is the time-dependent conductance of the $(i,j)$-th memristor during the update, and $t_0$ and $t_f$ define the the initial and final times respectively of the time interval over which the update process takes place for each device. In order to compute the energy dissipated of the $(i, j)-$th memristor during the update, $E_{\text{u}}^{ij}$, we use the trapezoidal rule to approximate the integral during the application of $V_{\text{u}}$ using $E_{\text{u}}^{ij} \approx \Delta t_{\text{u}} /2 \left( V_{\text{u}}^2(t_{k-1})G_{ij}(t_{k-1}) + V_{\text{u}}^2(t_{k})G_{ij}(t_{k})  \right)$, where we choose $\Delta t_{\text{u}} = 10$ ns and $G_{ij}(t_k/t_{k-1})$ is the conductance of the $(i,j)$-th memristor at the end/beginning of the update process. The total dissipated energy during one epoch is the sum over all devices $E_{\text{u}} = \sum_{ij} E_{\text{u}}^{ij}$. 

\section{Results}
To study the impact of the P/D curve parameters on the learning process performed in our simulated neural networks, we use a standard machine learning benchmark: pattern recognition on the MNIST dataset, which consists of 60000 images of 28 × 28 pixel digits from 0 to 9. We considered two network architectures: a simple perceptron (SP) with 28 × 28 input neurons  and 10 output neurons, and a deep neural network (DNN) with 28 × 28 input neurons, a hidden layer with 100 neurons, and an output layer with 10 neurons. In both cases, a softmax activation function was used at the output layer, and a ReLU function was applied to the hidden layer of the DNN. Multiclass cross-entropy was used as the loss function. \cite{Goodfellow2016}

\begin{figure}[H]
\centering
\includegraphics[width=1 \textwidth]{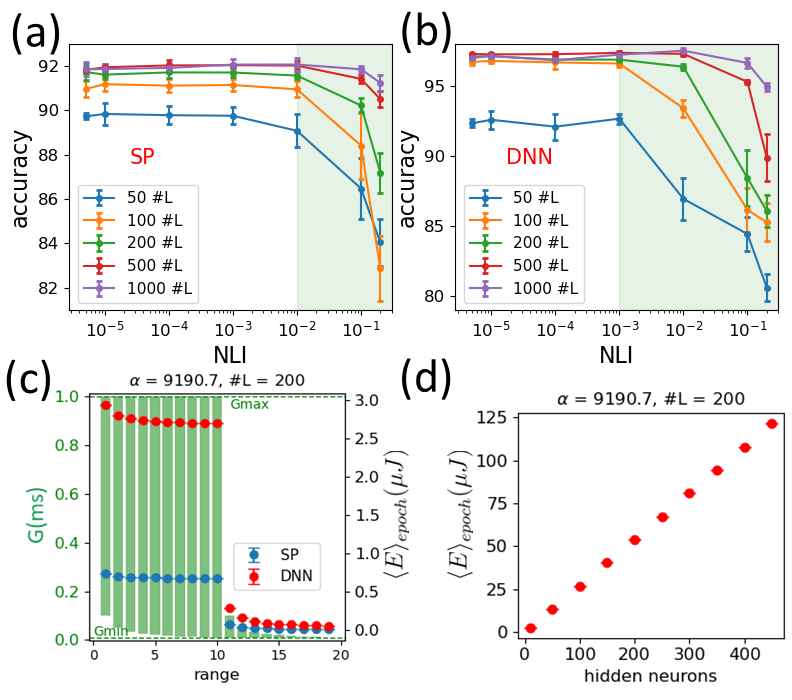}
\caption{Accuracy vs. NLI for (a) SP and (b) DNN and $\#$L spanning values of $50, 100, 200, 500$ and $1000$. The accuracy obtained remains almost constant for NLI $< 10^{-2}$ in the SP case and NLI $< 10^{-3}$ in the DNN case. For NLI values higher than this thresholds the accuracy starts to decrease (green region). The same results where obtained for different ranges of conductance. (c) Average energy consumption per epoch (right scale) for different conductance ranges (left scale) for SP (blue dots) and DNN (red dots). The parameters of the P/D curves used to calculate the power consumption were $\alpha$ = 9190.7 and  $\#$L = 200. (d) Scaling of the average energy consumption per epoch as a function of the number of neurons in the hidden layer.}
\label{Fig.3}
\end{figure}

To avoid overfitting during the learning process, we performed k-fold cross-validation \cite{Geron} with k=5, training on 4 folds and validating on the remaining one in each training realization. Each epoch consisted of processing all training samples divided into mini-batches of size 32.



The final accuracy after training vs NLI is shown in Fig. \ref{Fig.3}(a) for SP and in Fig.\ref{Fig.3}(b) for the DNN. Each curve in both figures corresponds to a different number of levels $\#L$ in the P/D curves.
For fixed $\#L$, we see that the accuracy remains basically unchanged up to NLI = $10^{-2}$ for SP and NLI = $10^{-3}$ for DNN, decreasing afterwards. This provides an estimate of how much nonlinearity in the P/D curves the crossbar array can tolerate before accuracy degrades. An improvement in accuracy is observed as $\#L$ increases. This makes sense since a higher density of levels brings the system closer to continuous updates, allowing for more precise convergence to the absolute minimum of the loss function. On the other hand, reducing the number of levels decreases the precision of updates, limiting the system's ability to find minima. From this we can conclude that the number of available levels in the P/D curves plays a role similar to an effective learning rate. 

The highest accuracy was achieved with the DNN for NLI $\le$ $10^{-3}$ and $\#L$ $\ge$ 100, reaching a value of $\approx$ 97\%, which is very close to the benchmark value of $\approx$ 98\% reported for an ideal network with exact synaptic weight actualization \cite{illing_2019}.
The same trend was observed for different conductance ranges for SP (Fig. SM-2) .

The estimated average energy consumption per epoch $\langle E \rangle_{epoch}$ is shown in Fig. \ref{Fig.3}(c) for both SP and DNN, using different conductance ranges between $G_{max} = 1$ mS and $G_{min} = 10$ mS, with parameters $\alpha$ = 9190.7 and $\#L$ = 200, in order to obtain linear P/D curves. From these results, we can see that for both SP and DNN, energy consumption is higher for conductance ranges where the highest values are closer to the maximum value $G_{min}$, which allows higher current values to be established in the crossbar. Conversely, as the conductance ranges move away from $G_{max}$ and closer to $G_{min}$, the currents decrease, resulting in lower energy consumption, which is expected since energy dissipation is linear on conductance. For all conductance ranges, the energy dissipated by the DNN is higher than that of the SP. This is expected, as the calculation involves summing the energy dissipation across all devices in the array. In all cases, we chose parameters assuring that accuracy takes its highest possible value (See Fig SM-3).


\begin{figure}[H]
\centering
\includegraphics[width=1 \textwidth]{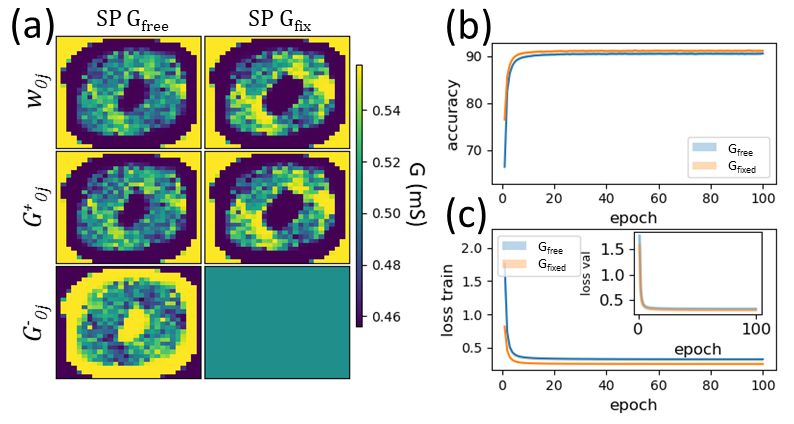}
\caption{ ((a), first row) Synaptic weight map for the first output neuron of SP ($w_{0j}$), using both $G_{free}$ (first column) and $G_{fix}$ (second column) update methods. ((a), middle row) Conductance map for $G_{0j}^+$ memristors. Each column shows results for the same neurons and layers as in the top row. ((a), bottom row) Conductance map for the $G_{0j}^-$ memristors; again, each column corresponds to the same neurons and layers as in the top row. (b) Accuracy evolution as a function of epochs. The solid line represents the mean values and the shaded band indicates the standard deviations. The blue line corresponds to the $G_{free}$ update method, while the orange line represents $G_{fix}$. (c) Training loss for both methods, with the validation loss shown in the inset. In both cases, convergence was achieved. A total of 200 training realizations were performed.}
\label{Fig.4}
\end{figure}

Finally, Fig. \ref{Fig.3}(d) demonstrates the average energy per epoch as a function of the number of neurons in the hidden layer. As anticipated, the energy increases linearly with the number of nodes, thereby revealing the direct influence of hidden-layer size on the overall energy consumption.

In Fig. \ref{Fig.4}(a) (first column), we plot the synaptic weight map ($w_{0j}$) together with the corresponding conductance maps ($G_{0j}^+$ and $G_{0j}^-$) for the SP, obtained after the training process and corresponding to the first output neuron.
These weight maps, $w_{mj}$, indicate how much each input neuron $j$ contributes to activating the output neuron $m$ (in this case, $m = 0$). For SP architectures, each output neuron is associated with a specific digit class, so the weight map can be interpreted as a template or prototype representative of that class.

In the synaptic weight map $w_{0j}$, a well defined template of the digit 0 (associated with the first output neuron) can be observed. Additionally, the synaptic weight values are highest at the periphery of the image and suddenly drop toward the center, reaching minimum values as the representative image of the class begins to take shape. 

The same pattern can be found in the conductance map $G_{0j}^+$, but interestingly, in the $G_{0j}^-$ map, the pattern appears as a 'negative image', that is, the maximum values become minimum and vice versa.
Since the conductances of both memristors in the differential pair are updated at each training step, it can be interpreted that the network stores the learned information redundantly, encoding the representative image of the class also in negative, through the $G_{0j}^-$ conductance values of the differential pair.
This suggests that the energy consumption during training can be reduced by eliminating this redundant learning. One way to achieve this is by fixing the $G_{ij}^-$ conductance to the midpoint of the conductance range defined by the P/D curves, and updating only the $G_{ij}^+$ values. We define this approach as $G_{fix}$, in contrast to $G_{free}$ where all conductances are adjusted during training. The obtained maps from this learning process, using linear P/D curves, are similar to the previous one (Fig. \ref{Fig.4} (second column)) except for $G_{ij}^-$ which remains constant in the fixed value. 
By examining the accuracy (Fig. \ref{Fig.4}(b)), training loss (Fig. \ref{Fig.4}(c)), and validation loss (Fig. \ref{Fig.4}(c, inset)) curves as a function of epochs, we observe convergence in both cases ($G_{free}$ in blue and $G_{fix}$ in orange). Two hundred learning realizations were performed; the solid lines represent the mean curves, and the shaded bands indicate the standard deviation. Statistically, the small difference between the results (0.6$\%$) suggests that both methods are equivalent in terms of learning performance on the crossbar.

\begin{figure}[H]
\centering
\includegraphics[width=1 \textwidth]{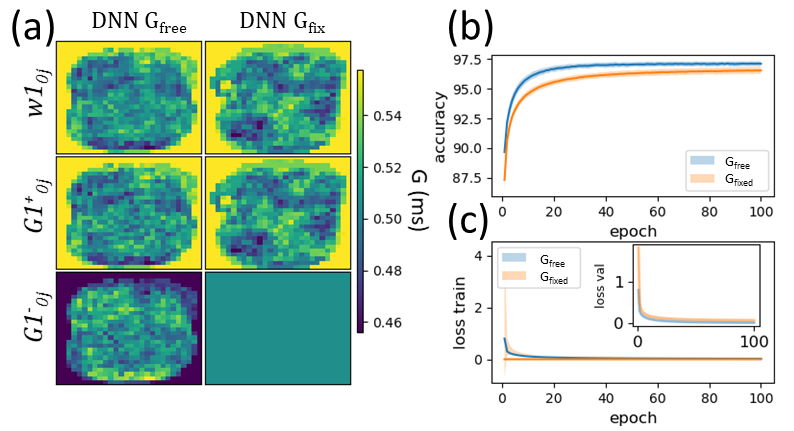}
\caption{((a) first row) Synaptic weight ($w1_{0j}$) map for the first neuron of the hidden layer of DNN, using $G_{free}$ update method (first column), and $G_{fix}$ (second column).  (Left, middle row) Conductance map for the $G1_{0j}^+$ memristors, each column shows results for the same neurons and layers as in the top row. (Left, bottom row) Conductance map for the $G1_{0j}^-$ memristors, again, each column corresponds to the same neurons and layers as in the top row. (b) Accuracy convergence as a function of epochs. The solid line represents the mean, and the shaded band indicates the standard deviation. The blue line corresponds to the $G_{free}$ update method, while the orange line represents $G_{fix}$. (c) Training loss for both methods, with the validation loss shown in the inset. In both cases, convergence was achieved. A total of 200 training realizations were performed. }
\label{Fig.5}
\end{figure}

For the hidden layers in a DNN, the interpretation of synaptic weight maps is not as straightforward as in the SP. 
In the case of the output layers, synaptic weight maps cannot be directly interpreted as class representatives due to the dimensional mismatch between the weight maps and the input images.
Nevertheless, the strategy  implemented in the SP case can still be applied to the synaptic weights of the DNN, by training $G^{+}_{ij}$ while fixing $G^{-}_{ij}$. This is illustrated in Fig.~\ref{Fig.5} (for the output layer, see Fig. SM-4). 
The evolution of accuracy, training loss, and validation loss under this approach is presented in Fig.~\ref{Fig.5}(a)--(c). 
We find that fixing $G^{-}_{ij}$ results in only a minor decrease in accuracy, less than $0.58\%$.

Having established that fixing one synaptic weight in each differential pair does not significantly affect the learning performance of either SP or DNN, we next estimate the associated average energy savings. In the G\textsubscript{fix} approach, only half of the memristors in the crossbar receive voltage pulses, directly reducing the energy required for training—the most time-consuming stage.

\begin{figure}[H]
\centering
\includegraphics[width=0.9 \textwidth]{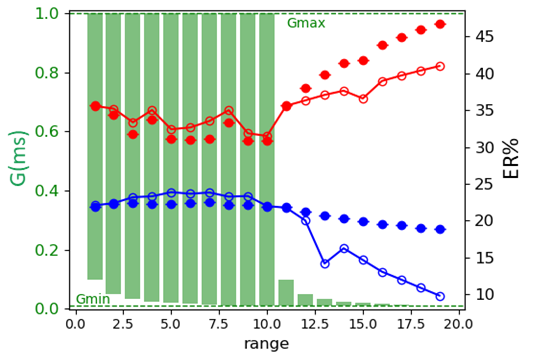}
\caption{Energy reduction (ER) for the $G_{fix}$ method compared to $G_{free}$, for different conductance ranges. Solid blue dots represent the SP results and red dots represent the DNN. Open dots correspond to the estimated values of $\langle E^p \rangle_{epoch}$. The solid lines are shown for visualization purposes. One hundred training realizations were performed for each point.   }
\label{Fig.6}
\end{figure}

We notice that writing events dominate the energy consumption, accounting for roughly 90$\%$ and 95$\%$ of the total energy in G\textsubscript{fix} and G\textsubscript{free}, respectively, as they adjust the synaptic weights and memristor conductivities. Fig.~\ref{Fig.6} summarizes the impact of the G\textsubscript{fix} strategy on the overall average energy consumption during training, both for SP and DNN. The figure reports the percentage of energy reduction (ER) with respect to the conventional approach, in which both devices are updated at every training step, for a variety of memristor conductance ranges. Blue points correspond to SP, while red points correspond to DNN.

For the SP, the energy reduction, defined as $ER = \left( 1 - E_{\mathrm{fix}}/E_{\mathrm{free}} \right) \times 100$, remains close to 20 $\%$ across the first set of conductance ranges (where $G_{max}$ = 1 mS and $G_{min}$ varies between 0.1 mS and 10 $\mu$S). When the conductance window is progressively narrowed (fixing $G_{min}$ at 10 $\mu$S and reducing $G_{max}$ from 0.1 mS to 11.1 $\mu$S), the ER slightly decreases to $\approx$ 15 $\%$ for the narrowest ranges. In contrast, the DNN consistently exhibits higher reductions, starting at 30–35 $\%$ for wide conductance ranges and increasing monotonically as the conductance window narrows and $G_{max}$ drops below 0.1 mS. For the narrowest ranges considered, the energy savings exceed 45 $\%$.


As mentioned earlier, the reduction in energy consumption stems from fixing half of the memristors in each differential pair. However, the saving remains below 50 $\%$ for both SP and DNN, reflecting differences in the conductance distributions between the free and fixed cases within the allowed conductance band. Indeed, Fig.~SM-5 (for $G_{\mathrm{MIN}} =$ 10 $\mu S$ and $G_{\mathrm{MAX}} =$ 100 $\mu S$) shows that a noticeable asymmetry develops in the fixed case, with a larger population of higher-conductance states---which are more energy demanding. This reduces the energy saving below the expectable 50 $\%$ threshold.

To understand the distinct trends in energy reduction for SP and DNN as the conductance window narrows, we recall, first, that the dominant contribution to the energy per epoch arises from the application of writing pulses during training. As shown in the Supplementary Material, this energy can be estimated as $\langle E^p \rangle_{epoch} \approx \Delta t V^2 \# D /2 \left[ \sum_k \langle G_{ij}(t_k) \rangle / \#E + \Delta G/2 \right]$ where $\Delta t$ is the time width of the writing pulse, $V$ the pulse amplitude, $\#D$ the number of devices in the layer, $\#E$ the number of training epochs, and $\Delta G$ the conductance step in the P/D curves (assuming, for simplicity, the linear case). The energy is dominated by the term $\sum_k \langle G_{ij}(t_k) \rangle$, which represents the sum of the average device conductance over the $k$ training epochs. Its evolution during training, which should ensure convergence and maximize accuracy, is directly linked to changes in the distribution of conductances $G_{ij}$ and ultimately determines the overall training cost.  

The observed $ER$ trends arise from a subtle competition between the decay rates of $E_{\mathrm{free}}$ and $E_{\mathrm{fix}}$, as shown in (Fig. SM-6). For SP (DNN), $E_{\mathrm{fix}}$ decreases with narrowing conductance windows at a higher (lower) rate than $E_{\mathrm{free}}$, thereby reducing (increasing) $ER$. These differences ultimately reflect distinct evolutions of the $G_{ij}$ distributions, as previously mentioned, in SP and DNN during training.  

Overall, Fig.~\ref{Fig.6} highlights that while the $G_{\mathrm{fix}}$ strategy reduces training energy in both SP and DNN architectures, its benefits are particularly pronounced in deeper networks. This makes DNNs especially well suited to exploit this approach, achieving savings of up to 45\% of the total training cost.

\section{Discussion and Conclusions}

Our simulations provide a comprehensive assessment of the trade-offs involved in training 
memristor-based neural networks under realistic device constraints. By systematically varying the 
linearity of potentiation/depression curves, the conductance window, and the number of available 
conductance levels, we identified the device parameters that most strongly influence learning 
accuracy and energy efficiency. We found that both SP and DNN architectures can tolerate a finite 
degree of nonlinearity in the P/D curves before accuracy begins to degrade, with DNNs requiring 
stricter linearity conditions (NLI $\leq 10^{-3}$) compared to SPs (NLI $\leq 10^{-2}$). Increasing the 
number of discrete conductance levels (\#L) effectively improves convergence, acting as a surrogate 
for a finer learning rate.

Nonetheless, our proposed \textit{$G_{fix}$} strategy---where one memristor of each differential pair is fixed---
demonstrates that energy consumption during training can be reduced by nearly 45\% in DNNs 
without compromising accuracy, indicating that careful co-optimization of device operating windows and 
training schemes can yield substantial efficiency gains.

Overall, our study underscores the potential of the Manhattan update rule as a hardware-friendly 
training scheme that balances learning performance with energy efficiency. The results highlight 
that future memristive hardware should not only focus on improving device-level characteristics 
(linearity, resolution, and stability), but also exploit algorithmic strategies such as redundant-weight 
fixing to minimize energy costs. These insights are particularly relevant for edge AI applications, 
where low-power and online learning capabilities are essential. We expect that combining realistic 
device engineering with training-rule co-design will pave the way toward scalable and energy-
efficient neuromorphic hardware.

\section*{SUPPLEMENTARY MATERIAL}

See the supplementary material for details on the approximations used to evaluate NN energy consumption, together with additional numerical results corresponding to both SP and DNN performance.

\section*{ACKNOWLEDGMENTS}

We acknowledge the support from EU-H2020-RISE project MELON (Grant No. 872631). We gratefully thank Beatriz Noheda and the University of Groningen for the access to their high performance computation facilities.

\section*{AUTHOR DECLARATIONS}

\textbf{Conflict of Interest}

The authors have no conflicts to disclose.

\textbf{Author Contributions}

W. Quiñonez led the investigation and data curation, and contributed to the formal analysis, conceptualization, and preparation of the original draft. M. J. Sánchez contributed to conceptualization, investigation, and funding acquisition, and participated in the manuscript review. D. Rubi led the conceptualization and funding acquisition, contributed to the investigation, and co-led the preparation and review of the manuscript.

\section*{DATA AVAILABILITY}
The data that support the findings of this study are available
from the corresponding author upon reasonable request.

\bibliography{references}

\end{document}